\documentclass[%
superscriptaddress,
preprint,
amsmath,amssymb,
aps,
pra,
noeprint
]{revtex4-2}

\usepackage{graphicx} % Required for inserting images
\usepackage[capitalise]{cleveref}
\graphicspath{{figs/}}

\usepackage{float}
\usepackage{siunitx}
\usepackage{booktabs}    % Professional horizontal rules (\toprule, \midrule, \bottomrule)
\usepackage{array}       % Enhanced column alignment control

\usepackage[defaultcolor=red]{changes}

\begin{document}

\title{Consistent Modeling of SEI Growth via Electron Diffusion Validated Across Diverse Cell Specifications and Operating Conditions in LFP/Graphite Cells}

\author{Lukas Köbbing}
\affiliation{Institute of Engineering Thermodynamics, German Aerospace Center (DLR), Wilhelm-Runge-Straße 10, 89081 Ulm, Germany}
\affiliation
{Helmholtz Institute Ulm (HIU), Helmholtzstraße 11, 89081 Ulm, Germany}

\author{William Black}
\affiliation{Department of Chemistry, Dalhousie University, Halifax B3H 4R2, Canada}

\author{Claire Floras}
\affiliation{Department of Mechanical Engineering, Dalhousie University, Halifax B3H 4R2, Canada}

\author{Saad Azam}
\affiliation{Department of Chemistry, Dalhousie University, Halifax B3H 4R2, Canada}

\author{Eniko Zsoldos}
\affiliation{Department of Chemistry, Dalhousie University, Halifax B3H 4R2, Canada}

\author{Ben Tang}
\affiliation{Department of Physics and Atmospheric Science, Dalhousie University, Halifax B3H 4R2, Canada}

\author{J. R. Dahn}
\affiliation{Department of Chemistry, Dalhousie University, Halifax B3H 4R2, Canada}
\affiliation{Department of Mechanical Engineering, Dalhousie University, Halifax B3H 4R2, Canada}
\affiliation{Department of Physics and Atmospheric Science, Dalhousie University, Halifax B3H 4R2, Canada}

\author{Birger Horstmann}
\email{birger.horstmann@dlr.de}
\affiliation{Institute of Engineering Thermodynamics, German Aerospace Center (DLR), Wilhelm-Runge-Straße 10, 89081 Ulm, Germany}
\affiliation
{Helmholtz Institute Ulm (HIU), Helmholtzstraße 11, 89081 Ulm, Germany}
\affiliation{Faculty of Natural Sciences, Ulm University, Albert-Einstein-Allee 47, 89081 Ulm, Germany}

\date{\today}

\begin{abstract}
    Understanding and predicting degradation in lithium-ion batteries is critical for their widespread deployment. The continuous growth of the solid-electrolyte interphase (SEI) remains a dominant aging pathway, yet accurately modeling its kinetics across diverse cell specifications and operating conditions remains a significant challenge. Here, we present and validate a physically grounded SEI growth model mediated by the electron diffusion mechanism. This framework intrinsically captures the state of charge (SOC) dependence of degradation through the impact of the anode potential on the electron concentration, which governs the rate of SEI growth. The presented model explicitly accounts for variations in electrolyte salt, anode material, temperature, and cycling rate. Validation against experimental aging data for LFP/graphite cells demonstrates high predictive accuracy across a broad range of cell specifications and cycling conditions, including confined SOC windows, electrolyte salts (LiFSI, LiPF$_6$), artificial graphites (AGC, AGD), temperatures ($40-70\,^\circ\mathrm{C}$), and C-rates (C/20 $–$ 1C). Remarkably, a single, consistent set of parameters is applied to simulate the capacity fade for all experiments without recalibration. While quantitative overestimations occur during high-SOC storage periods, the framework consistently captures the observed degradation trends. Our findings underscore that lower SOC operation mitigates degradation, establishing a physics-based model for predicting long-term lithium-ion battery aging.

    \ \\
    \textbf{Keywords}: solid-electrolyte interphase (SEI), electron diffusion, battery degradation modeling, state of charge dependence, LFP/graphite batteries, physics-based aging prediction
\end{abstract}

\maketitle

\section{Introduction}

Lithium-ion batteries (LIBs) have established an indispensable role in modern energy storage applications \cite{Winter2018, Li2018}. Within this landscape, lithium iron phosphate (LFP) chemistry is gaining increasing relevance due to its favorable cost and material abundance \cite{Chen2024}. However, regardless of the specific chemistry, battery degradation manifests through a variety of mechanisms \cite{Hu2024, Han2019, Edge2021, OKane2022, Birkl2017}. Consequently, understanding and mitigating these degradation pathways remains a critical challenge for virtually all battery deployments, particularly for LFP-based systems, which are increasingly utilized as large-scale stationary energy storage solutions.

Among the various degradation mechanisms, the continuous growth of the solid-electrolyte interphase (SEI) is widely recognized as the dominant aging pathway \cite{Peled2017, Verma2010, Bloom2010}. During storage, SEI formation typically follows a square-root-of-time dependence \cite{Attia2020, Broussely2001, Smith2011, Wang2014, Kolzenberg2022} and exhibits a pronounced dependence on the state of charge (SOC), as evidenced by experimental studies \cite{Keil2016, Keil2017, Naumann2018}. Building on these observations, we previously introduced and evaluated a physically grounded SEI growth model based on electron diffusion, which accurately captures the SOC-dependent capacity fade during storage periods \cite{Single2018, Koebbing2023Growth}. Crucially, our comparative analysis demonstrated the superiority of the electron diffusion mechanism over traditional solvent-diffusion-mediated models \cite{Koebbing2023Growth}. The electron diffusion mechanism has since received independent experimental validation through cyclic voltammetry measurements \cite{Krauss2024}. Beyond storage conditions, studies on SEI evolution during cycling revealed that electron diffusion dominates the SEI growth over electron migration until late stages of battery aging \cite{Kolzenberg2020, Kolzenberg2022}. Notably, recent experimental findings continue to underscore a strong SOC dependence during cycling, particularly when operation is confined to specific SOC windows~\cite{Zsoldos2024}. The investigation was supported by the observation of larger heat flows due to SEI reactions at the graphite anode at higher SOC, implying a higher reactivity of the anode at lower potentials \cite{Zsoldos2024}.

In addition to the SOC dependence, SEI growth is critically sensitive to cell design and operating conditions. The selection of anode materials (e.g., a specific artificial graphite), electrolyte composition (including solvent systems, salts, and functional additives), and operating temperature profoundly influence degradation rates \cite{Black2025, Logan2020, Azam2025}. At extended timescales and elevated temperatures, iron dissolution from the cathode can become significant, leading to accelerated, non-square-root SEI growth profiles \cite{Black2025}. However, for cells cycled under moderate conditions, such non-linear contributions remain negligible at early and intermediate aging periods and are expected to emerge only at advanced states of aging \cite{Black2025}.

In this paper, we present our physically derived SEI growth model mediated by the electron diffusion mechanism and compare the model predictions to experimental data across a broad spectrum of cell specifications and operating conditions. Our work was significantly motivated by the systematic experimental data showing clear SOC-dependence during cycling presented in Ref.~\cite{Zsoldos2024}, which we herein analyze and explain using our SEI growth model. Therefore, this work distinguishes from our previous SEI growth modeling publications \cite{Single2018, Koebbing2023Growth, Kolzenberg2020, Kolzenberg2022} in modeling and analyzing systematic cycling data in confined SOC regions and including variations in electrolyte salt, anode material, temperature, and cycling rate. By analyzing both published datasets and newly generated experimental results, we demonstrate that our model reproduces the observed capacity fade under diverse cycling protocols, temperatures, electrolyte salts, and artificial graphites. Our model remains simple in use and highly adaptable, allowing straightforward parameterization for new cell configurations. The remainder of this paper is structured as follows: Section \ref{sec:theory} details the mathematical formulation of our model and the parameterization variables. Section \ref{sec:experiments} outlines the experimental cell specifications as well as the testing protocols. Section \ref{sec:results} presents a comprehensive validation of the model against experimental aging data for a wide range of cycling in confined SOC windows for different electrolyte salts, artificial graphites, and temperatures, as well as different cycling protocols. Finally, Section \ref{sec:conclusion} summarizes the key findings and outlines directions for future research.

\section{Theory}
\label{sec:theory}

The electron diffusion model for solid-electrolyte interphase (SEI) growth originally presented in Ref.~\cite{Single2018} was subsequently extended to account for cycling conditions in Ref.~\cite{Kolzenberg2020} and validated in Refs.~\cite{Koebbing2023Growth, Kolzenberg2022}. This framework is grounded in the chemical equilibrium of electrons at the anode/SEI interface, followed by the diffusion of electrons through the SEI toward the SEI/electrolyte interface due to an electron concentration gradient inside the SEI. At the SEI/electrolyte interface, the electrons participate in a parasitic reaction, generating additional SEI and thereby reducing the cell capacity. While electron migration due to a potential gradient inside the SEI can also contribute to charge transport, it has been shown to become significant only at later stages of aging \cite{Kolzenberg2020} and is therefore neglected here. Nevertheless, our model explicitly incorporates the anode overpotential generated during cycling, as it critically influences the equilibrium of electrons at the anode/SEI interface and, consequently, the electron-diffusion-mediated SEI growth.

The rate of SEI formation according to the electron diffusion model, expressed as the time derivative of the capacity $Q_\mathrm{SEI}$ consumed by the SEI, is given by \cite{Single2018, Kolzenberg2020, Koebbing2023Growth}
\begin{equation}
    \partial_t Q_\mathrm{SEI} = \xi \exp \left(-\frac{FU}{RT}\right) \frac{1}{Q_\mathrm{SEI}+Q_\mathrm{SEI,0}} .
    \label{eq:SEI-model-simplest}
\end{equation}
Here, $\xi$ is a prefactor that encapsulates all dependencies of the SEI growth on the cell specifications and environmental conditions, such as electrolyte composition, anode material, and operating temperature $T$. In \cref{eq:SEI-model-simplest}, $F$ denotes the Faraday constant and $R$ is the universal gas constant, while $Q_\mathrm{SEI,0}$ represents the initial SEI capacity. The anode potential $U$ is determined by the open-circuit voltage (OCV), $U_0(SOC)$, which depends on the state of charge (SOC), and the anode overpotential $\eta$.

To systematically account for variations in the used electrolyte salt and graphite type in experiments, we reformulate the SEI growth from \cref{eq:SEI-model-simplest} by decomposing the prefactor $\xi$ into multiplicative components, separately describing the impact of changing these specifications.
Moreover, we explicitly account for the operating temperature $T$, the $SOC$, and the anode overpotential $\eta$.
The rewritten model, explicitly stating these influences, defines the normalized SEI capacity $\tilde{Q}_\mathrm{SEI}$, which is normalized by the initial cell capacity $Q_\mathrm{cell}$, as
\begin{equation}
    \partial_t \tilde{Q}_\mathrm{SEI} = \xi_\mathrm{ref} \xi_\mathrm{electrolyte} \xi_\mathrm{AG} \frac{\exp \left(-\frac{E_\mathrm{A}}{kT}\right)}{\exp \left(-\frac{E_\mathrm{A}}{kT_\mathrm{ref}}\right)}  \frac{\exp \left(-\frac{F(U_0(SOC)-\eta)}{RT}\right)}{\tilde{Q}_\mathrm{SEI}+\tilde{Q}_\mathrm{SEI,0}} .
    \label{eq:SEI-model-used}
\end{equation}
In this formulation, $\xi_\mathrm{ref}$ denotes the baseline parameter for a reference cell configuration at reference temperature $T_\mathrm{ref}$. The dimensionless factors $\xi_\mathrm{electrolyte}$ and $\xi_\mathrm{AG}$ scale the SEI growth kinetics in response to variations in the used electrolyte or artificial graphite, respectively, and are normalized to unity for the reference case. Note that $\xi_\mathrm{electrolyte}$ in general covers changes in electrolyte salt, solvent, and additives. However, variations in the solvent and the additives are not investigated in this study. Furthermore, the Arrhenius-type term $\exp(-E_\mathrm{A}/(kT))$ introduces an explicit temperature dependence, where $E_\mathrm{A}$ represents the activation energy of the SEI formation process, namely the limiting electron diffusion step. As experiments start with fresh cells right after formation, we assume the normalized initial SEI capacity relevant for the electron diffusion process to be negligible in all simulations, $\tilde{Q}_\mathrm{SEI,0}=\num{1e-6}$. 

As highlighted in \cref{eq:SEI-model-used}, the anodic overpotential $\eta$ plays a crucial role in modulating the SEI growth rate under cycling conditions. We compute the anode overpotential $\eta$ from the Butler--Volmer equation
\begin{equation}
    \eta = \pm \frac{2RT}{F} \mathrm{asinh} \left(\frac{C_\mathrm{rate}}{j_0 \xi_\mathrm{AG}} \frac{\exp(-E_\eta/(kT_\mathrm{ref}))}{\exp(-E_\eta/(kT))}\right) .
    \label{eq:eta}
\end{equation}
Here, the sign accounts for charging ($+$) and discharging ($-$) currents. During lithiation (charging), $\eta > 0$ lowers the anode potential, increasing the driving force of the electron diffusion mechanism. The parameter $C_\mathrm{rate}$ is the applied C-rate, while $j_0$ serves as a fitting parameter proportional to the exchange current density. The activation energy $E_\eta$ characterizes the temperature dependence of the overpotential, and $\xi_\mathrm{AG}$ again accounts for the impact of the graphite used. The parameter $j_0$ is calculated using the estimated overpotential $\eta_\mathrm{ref}$ at a reference temperature of \SI{40}{\degreeCelsius}.

Collectively, \cref{eq:SEI-model-used} and \cref{eq:eta} provide a compact yet comprehensive description of electron-diffusion-mediated SEI growth that explicitly incorporates the influence of key cell parameters and operating conditions. For a reference configuration, the model requires only a single fitting parameter, $\xi_\mathrm{ref}$, alongside an estimate or measurement of the anode overpotential $\eta_\mathrm{ref}$. When extending the framework to varying electrolyte salts or graphite types, we introduce the scaling factors $\xi_\mathrm{electrolyte}$ and $\xi_\mathrm{AG}$. The influence of the temperature is handled through the activation energies $E_\mathrm{A}$ for the electron diffusion and $E_\eta$ for the anode overpotential. With this parameterization, comprising six parameters, we conduct systematic simulations in Sec.~\ref{sec:results} to reproduce a broad spectrum of experimental aging data, encompassing variations in cycling SOC windows, the used electrolyte salt, artificial graphite material, operating temperature, cycling rate, and combined cycling-storage protocols.

\section{Experiments}
\label{sec:experiments}

\textbf{Pouch cells} – This work utilizes machine-made 402035-sized pouch cells (LiFUN Technologies, China) with an LFP cathode coupled to one of two different artificial graphite materials labeled AGC or AGD. The labeling of these graphite negative electrodes is consistent with previously published papers in our group \cite{Eldesoky2021, Eldesoky2022, Azam2025Improving}. The LFP pouch cells had a \SI{240}{mAh} nominal capacity and were balanced to \SI{3.65}{V} with an N:P ratio of 1.1. All pouch cells were received dry and then subsequently cut open in an argon (Ar) filled glovebox (GB) where they were vacuum baked at \SI{120}{\degreeCelsius} for \SI{14}{h} prior to being filled with electrolyte to remove residual moisture.

\textbf{Electrolyte preparation and pouch cell formation} – A base electrolyte solution of 1.5 M LiFSI or LiPF$_6$ in solvent mixtures of ethylene carbonate (EC) with dimethyl carbonate (DMC) and/or ethyl methyl carbonate (EMC) in weight ratios of 15(EC):85(DMC) or 30(EC):70(DMC) or 15(EC):15(EMC):70(DMC) was used in this study. Vinylene carbonate (VC) was used as an additive at various weight percentages either 2 wt.\,\% or 4 wt.\,\%. Electrolytes were prepared in an Ar-filled glovebox and stored for one day prior to pouch cell filling. LFP pouch cells (\SI{240}{mAh}) were filled with \SI{0.85}{\milli\liter} or \SI{1.0}{\milli\liter} of electrolyte. After filling, cells were vacuum sealed under \SI{-90}{kPa} gauge pressure in a vacuum sealer (MSK-115A, MTI Corp.). Sealing bars were set to \SI{165}{\degreeCelsius} and held in place for \SI{4}{s}. Subsequently, the cells were then transferred to \SI{40.0\pm 0.1}{\degreeCelsius} temperature-controlled chambers where they were charged to \SI{1.5}{V} and maintained at that voltage for \SI{16}{h} to ensure that the electrodes were sufficiently 'wetted' with electrolyte. After the wetting step, the cells underwent “formation” at the same temperature using a Maccor 4000 series battery tester. The formation step consisted of one charge-discharge cycle from \SI{2.5}{V} to \SI{3.65}{V} at a C/20 rate and one final charge to 50\% SOC (\SI{\sim3.3}{V}) by charging at C/10 for \SI{5}{h}. Following formation, cells were then taken inside the Ar-filled glove box, cut open, and re-sealed following the same procedure described above to remove gases generated during the formation process.

\textbf{Long-term cycling protocol} – Long-term cycling of the pouch cells was conducted in insulated temperature boxes set to \SI{40}{\degreeCelsius}, \SI{55}{\degreeCelsius}, or \SI{70}{\degreeCelsius} (\SI{\pm 1}{\degreeCelsius}) using a Neware Battery Testing System (Shenzhen, China). Pouch cells were cycled following a variety of protocols.

In cycling experiments and cycling rate experiments, cells were cycled from \SI{2.5}{V} to \SI{3.65}{V} at a rate of 1C, C/3, or C/20 in constant current constant voltage mode during charge with a C/20 cut-off current and at 1C, C/3, or C/20 in constant current mode during discharge. Every 50 cycles, a “check-up” cycle was included at C/20 (for cells cycling at C/3 or 1C) to probe low-rate capacity.

In SOC experiments, cells were cycled at a rate of C/3 using a constant current charge step which was limited by time and a constant current discharge limited by voltage for SOC ranges from 0\% to 25\%, 60\%, 80\%, or 100\% SOC. For cycling at the highest SOC range 75\% to 100\% again a C/3 rate was used, however, a constant current constant voltage charge and constant current discharge limited by time was used. In all cases, checkup cycles were completed every 200 cycles to determine full cell capacity \cite{Zsoldos2024}.

\begin{figure*}[htbp]
    % Figure #
    % width = 2 columns
    \centering
    \includegraphics[height=0.35\textwidth]{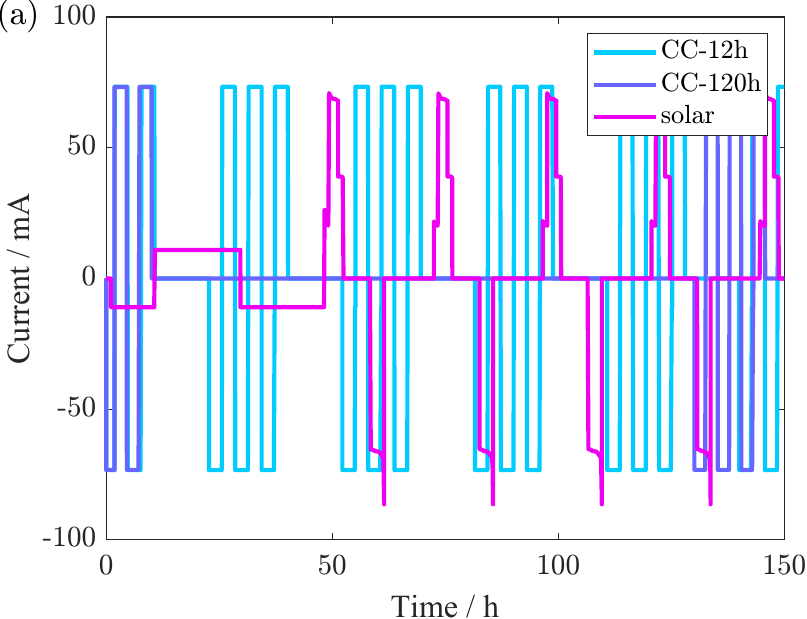}
    \hspace{0.5cm}
    \includegraphics[height=0.35\textwidth]{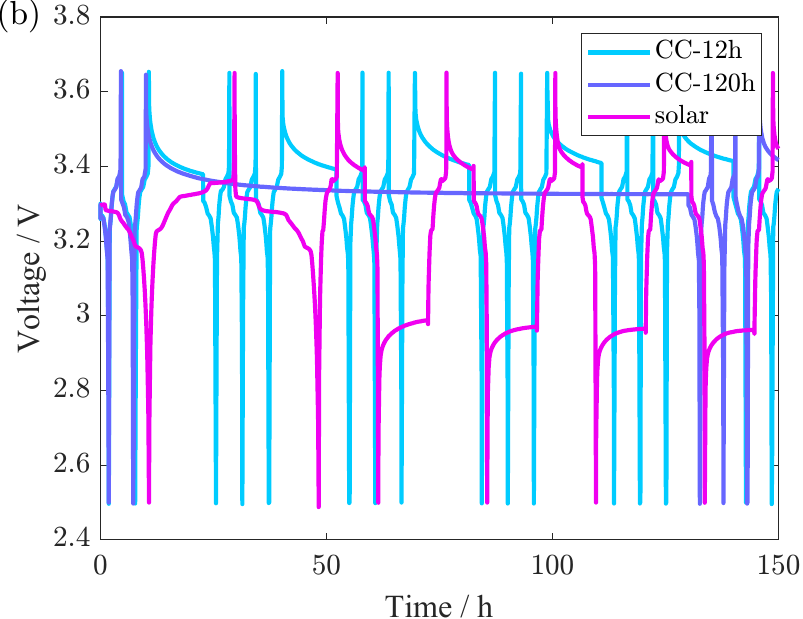}
    \caption{\textbf{Protocol details for mixed cycle-storage protocols.} The figure shows (a) current versus time and (b) voltage versus time for the protocols involving cycling and storage periods for \SI{12}{h} and \SI{120}{h} at 100\% SOC as well as the solar duty cycle.}
    \label{fig:protocols-info}
\end{figure*}

In cycle-storage experiments, cells were subject to both cycling and open-circuit voltage conditions. In these tests, cells were first cycled from \SI{2.5}{V} to \SI{3.65}{V} at C/3 using a constant current and a voltage-hold step where cells were held at \SI{3.65}{V} for \SI{1}{h} before discharge in which a constant current constant voltage discharge with a C/20 cut off current was used. Cells completed two full cycles between open-circuit stage periods, which were initiated after a voltage-hold step for \SI{1}{h} at \SI{3.65}{V}. Open-circuit storage periods varied in length with \SI{12}{h} or \SI{120}{h} storage periods. The current and voltage profiles during the cycle-storage protocols are visualized in \cref{fig:protocols-info}.

In solar duty experiments, charge and discharge profiles were developed to provide constant power to a cell used in a “\SI{24}{h} solar duty application” \cite{Ye2026}. The resulting protocol consisted of a \SI{4}{h} charging period broken into three stages: an initial \SI{1}{h} phase at roughly \SI{\sim 20}{mA}, a \SI{2}{h} phase at \SI{\sim 70}{mA}, and a final \SI{1}{h} phase at \SI{\sim 40}{mA} to the UCV at \SI{3.65}{V}. Following the completion of the charging protocol, each cell is held at a 100\% state of charge (SOC) for a \SI{6}{h} OCV storage period. The discharging phase down to a 0\% SOC was completed over an approximately \SI{3}{h} duration, at \SI{\sim 66}{mA} to the LCV \SI{2.5}{V}. To complete the 24-hour cycle, the cell was then rested at a 0\% SOC for \SI{11}{h}. The sequence then repeats. The current and voltage profiles during the solar duty protocol are visualized in \cref{fig:protocols-info}.

% plot current and voltage of protocols

\section{Results \& Discussion}
\label{sec:results}

In this section, we evaluate the predictions of our SEI growth model based on the electron diffusion mechanism described by \cref{eq:SEI-model-used} and \cref{eq:eta} against a comprehensive set of experimental data spanning various cell specifications and operating conditions. The results are presented systematically, beginning with simulations of a reference cell configuration operated within confined SOC windows in Sec. \ref{sec:SOC-windows}. We subsequently investigate the influence of varying electrolyte salts and artificial graphite types in Sec. \ref{sec:materials}, followed by an analysis of temperature effects in Sec. \ref{sec:temperatures} and different C-rates in Sec. \ref{sec:rates}. Finally, we examine mixed cycling and storage aging protocols in Sec. \ref{sec:storage-periods}. In all simulations, we consider the specific cycling protocols and conditions, especially accounting for the intermediately performed check-up cycles. We show all capacity data normalized to the initial cell capacity, which we determine as 98\% of the formation cycle discharge capacity or the first full-cycle discharge capacity under the investigated conditions.

Crucially, a single, consistent set of simulation parameters is employed to describe the entire range of experiments without recalibration for each condition. We apply the following parametrization strategy. First, $\xi_\mathrm{ref}$ is obtained by minimizing the deviation between simulation and experimental capacity fade for the reference case in confined SOC windows as discussed in Sec.~\ref{sec:SOC-windows}, assuming a constant anode overpotential $\eta_\mathrm{ref}$. Afterwards, $\xi_\mathrm{electrolyte}$ and $\xi_\mathrm{AG}$ are fitted to experiments varying the used salt and graphite, respectively, in Sec.~\ref{sec:materials}. The activation energy of the overpotential $E_\eta$ is determined from half-cell voltage polarization data for AGD as shown in the Supporting Information in Fig.~S2. The activation energy $E_\mathrm{A}$ is determined to match the experimental capacity fade at $\SIrange{40}{70}{\degreeCelsius}$ in Sec.~\ref{sec:temperatures}. The consistent set of parameters is summarized in \cref{tab:fit-parameters}.

\begin{table}[tbp]
    \centering
    \renewcommand{\arraystretch}{1.35}
    \begin{tabular}{l c c}
        \toprule
        \textbf{Parameter} & \multicolumn{2}{c}{\textbf{Value}} \\
        \midrule
        \multicolumn{3}{@{} l}{\textbf{Reference Configuration}} \\
        \multicolumn{3}{@{} l}{AGC, LiFSI, C/3, $T_\mathrm{ref}=\SI{40}{\degreeCelsius}$} \\
        \midrule
        $\xi_\mathrm{ref}$ & \multicolumn{2}{c}{\SI{2.2e-5}{h^{-1}} (fitted)} \\
        $\eta_\mathrm{ref}$ & \multicolumn{2}{c}{\SI{46}{mV} (estimated)} \\
        \midrule
        \multicolumn{3}{@{} l}{\textbf{Material Scaling Factors}} \\
        \midrule
        Graphite factor $\xi_\mathrm{AG}$ & AGC & 1 (fixed) \\
        & AGD & \SI{2.0}{} (fitted) \\
        Salt factor $\xi_\mathrm{electrolyte}$ & LiFSI & 1 (fixed) \\
        & LiPF$_6$ & \SI{1.6}{} (fitted) \\
        \midrule
        \multicolumn{3}{@{} l}{\textbf{Activation energies}} \\
        \midrule
        $E_\mathrm{A}$ (electron diffusion) & \multicolumn{2}{c}{\SI{0.4}{eV} (fitted)} \\
        $E_\mathrm{\eta}$ (overpotential) & \multicolumn{2}{c}{\SI{0.273}{eV} (fitted to $\eta_\mathrm{AGD}$)} \\
        \midrule
        \multicolumn{3}{@{} l}{\textbf{Fundamental Constants}} \\
        \midrule
        Faraday constant $F$ & \multicolumn{2}{c}{\SI{96485}{\C\per\mol}} \\
        Gas constant $R$ & \multicolumn{2}{c}{\SI{8.314}{\J \per\mol\per\K}} \\
        Boltzmann constant $k$ & \multicolumn{2}{c}{\SI{1.38e-23}{\J \per\K}} \\
        \bottomrule
    \end{tabular}
    \caption{Summary of the consistent set of model parameters.}
    \label{tab:fit-parameters}
\end{table}

\subsection{Influence of SOC windows of operation}
\label{sec:SOC-windows}

In this section, we validate our SEI growth model by comparing the simulation results with experimental data for full-cell capacity decay of LFP/graphite cells obtained from Ref.~\cite{Zsoldos2024}. This dataset contains cycling in various confined SOC windows. For the analysis of the impact of SOC windows, we consider a reference cell configuration comprising artificial graphite AGC, electrolyte salt LiFSI, and an operating temperature of \SI{40}{\degreeCelsius}. 

Regarding our model, as defined in \cref{eq:SEI-model-used}, the state of charge (SOC) determines the anode open-circuit voltage $U_0$, which intrinsically governs the electron diffusion rate and, consequently, the rate of SEI growth. Crucially, the influence of the SOC-dependent voltage is a physical feature of our model that requires no additional fitting parameters. For the simulations, we fit the OCV of the graphite anode to half-cell data with high precision as shown in the Supporting Information in Fig.~S1. We assume the anode overpotential $\eta$ is independent of SOC with a constant value estimated as $\eta_\mathrm{ref} = \SI{46}{mV}$ for this reference case as discussed in the Supporting Information in Sec.~SII. Consequently, the fitting procedure involves only a single parameter, $\xi_{\mathrm{ref}}$, while the material-specific factors $\xi_{\mathrm{AGC}}$ and $\xi_{\mathrm{LiFSI}}$ are normalized to unity.

\begin{figure*}[htbp]
    % Figure #
    % width = 2 columns
    \centering
    \includegraphics[height=0.35\textwidth]{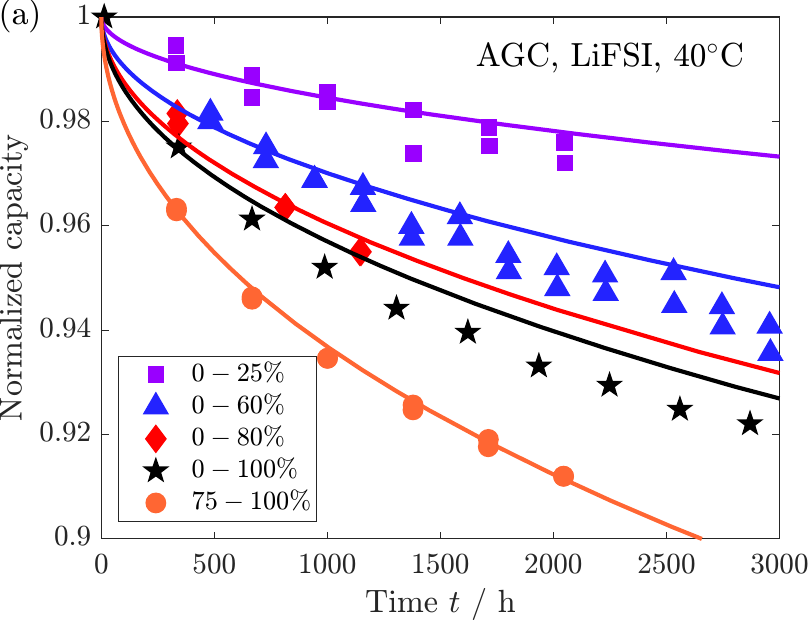}
    \hspace{0.5cm}
    \includegraphics[height=0.35\textwidth]{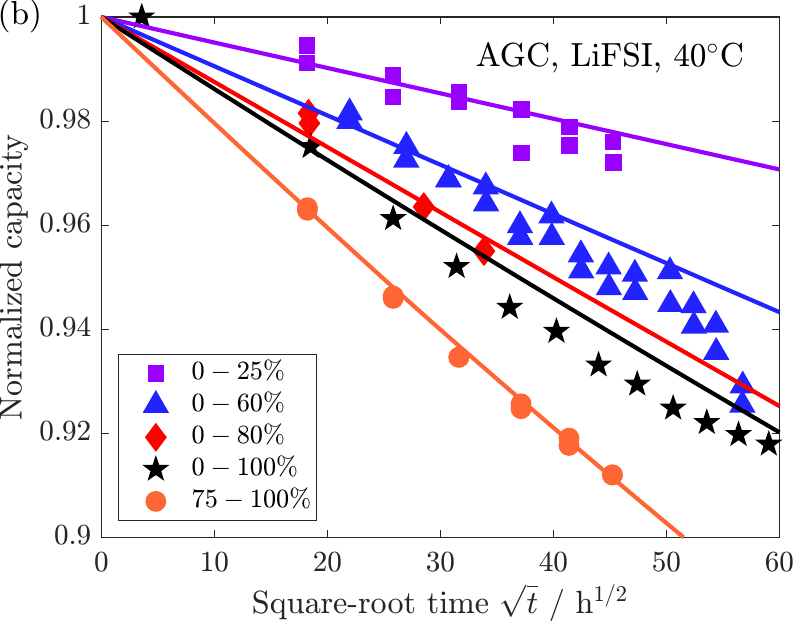}
    \caption{\textbf{SOC dependence.}
    Evolution of normalized capacity due to SEI growth for a reference cell configuration (artificial graphite AGC, LiFSI electrolyte) cycled at \SI{40}{\degreeCelsius} within different confined SOC windows. (a) Capacity evolution as a function of time. Solid lines denote simulation results, while markers indicate experimental measurements. (b) The same data plotted against the square root of time, demonstrating the linear trend characteristic of diffusion-limited SEI growth.
    Cell IDs: 489406-07, 491251-52, 491208-09, 487689, 489402-03.}
    \label{fig:capacity-LiFSI-40}
\end{figure*}

Figure~\ref{fig:capacity-LiFSI-40}(a) illustrates the remaining normalized capacity versus time, revealing accurate agreement between our simulations (solid lines) and experimental data (markers). The results demonstrate that cycling within high SOC regions (e.g., 75\%--100\%) leads to significantly greater capacity loss compared to low SOC regions (e.g., 0\%--25\%). This behavior is directly attributable to the anode potential. Lower anode potentials at high SOC accelerate SEI formation, whereas higher potentials at low SOC suppress it. Our simulation captures the experimental trend precisely, indicating that the intrinsic physics of the electron diffusion mechanism successfully reproduces the experimental degradation without further tuning.

Furthermore, \cref{fig:capacity-LiFSI-40}(b) presents the same data plotted against the square root of time. Both the simulation and the experiment reveal a linear relationship in this plot, confirming the characteristic square-root-of-time scaling typical of diffusion-limited SEI growth. 
This validates our assumption that the impact of electron migration on SEI growth is negligible.
In summary, this section demonstrates that our physically derived electron-diffusion-mediated SEI growth model accurately describes the SOC and time dependencies observed during cycling under reference conditions using only one fitting parameter, namely $\xi_{\mathrm{ref}}$. In the following steps, we extend our analysis to investigate variations in cell material specifications, specifically the electrolyte salt and artificial graphite used.

% plpt AGC, 40°C, LIFSI (2x)

\subsection{Influence of material choice}
\label{sec:materials}

The specific material choice of a battery cell, particularly the choice of electrolyte salt and artificial graphite, significantly influences the rate and amount of SEI growth. The electrolyte salt determines the chemical composition of the SEI layer formed, while the graphite morphology affects the active surface area. We investigate these variations by comparing LiPF$_6$-based electrolytes and a different artificial graphite (AGD) against our reference system with AGC and LiFSI.

\begin{figure*}[htbp]
    % Figure #
    % width = 2 columns
    \centering
    \includegraphics[height=0.35\textwidth]{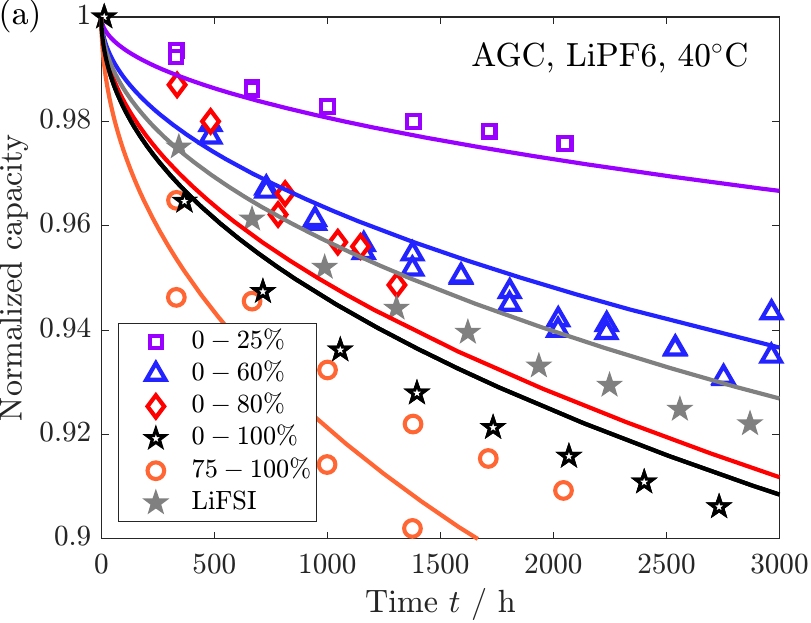}
    \hspace{0.5cm}
    \includegraphics[height=0.35\textwidth]{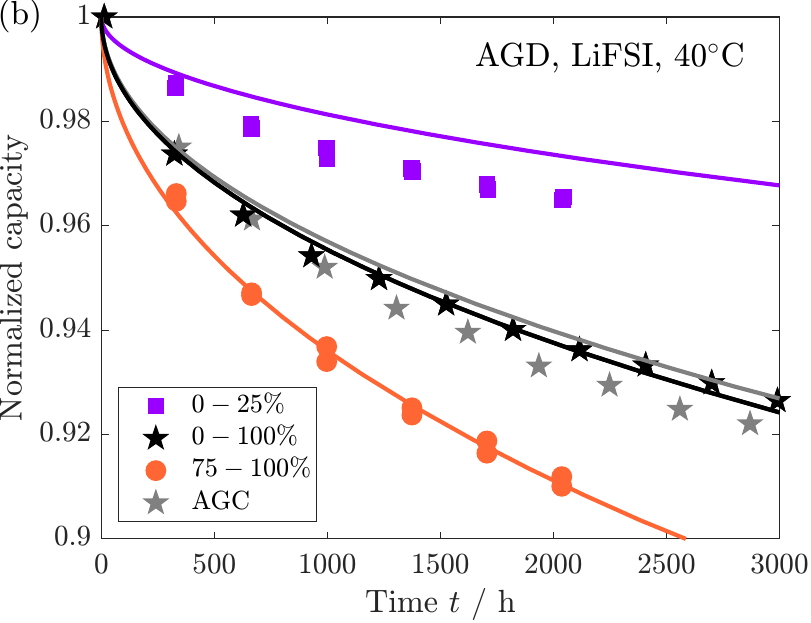}
    \caption{\textbf{Material dependence.} 
    Evolution of normalized capacity due to SEI growth at \SI{40}{\degreeCelsius} for (a) a different electrolyte salt, LiPF$_6$, and (b) a different artificial graphite material, AGD, in comparison to the reference configuration 0\%--100\% SOC data depicted in grey. The LiPF$_6$ cells were cycled within the same confined SOC windows as in \cref{fig:capacity-LiFSI-40}, but some SOC windows are missing for AGD. 
    Lines denote simulation results, while markers indicate experimental measurements.
    Cell IDs: (a) 489408-09, 491253-54, 491210-11, 488180, 489404-05. (b) 489390-91, 487668, 489388-89.}
    \label{fig:capacity-material-choices}
\end{figure*}

First, we examine the influence of the electrolyte salt, while keeping the artificial graphite AGC and the temperature $T_\mathrm{ref}=\SI{40}{\degreeCelsius}$ of the reference case. Figure \ref{fig:capacity-material-choices}(a) illustrates the capacity loss for cells utilizing LiPF$_6$ in simulation (lines) and experiment (markers) for cycling in various SOC windows. Compared to the reference case with LiFSI presented in \cref{fig:capacity-LiFSI-40}(a), the qualitative trend of the impact of the SOC ranges remains consistent also for LiPF$_6$. Cycling within high-SOC regions induces significantly more severe degradation than low-SOC windows (e.g., 75\%--100\% versus 0\%--25\%). For comparison with the reference case, we have included the simulation and data for the reference case using LiFSI for cycling in the 0\%--100\% SOC region in grey.
This comparison reveals that the capacity loss with LiPF$_6$ is higher than that with LiFSI.

In the experiment, the electrolyte salt dictates the reaction kinetics through the generation of specific SEI species. In our model, we account for this by adjusting the salt-dependent prefactor $\xi_\mathrm{electrolyte}$. By tuning $\xi_\mathrm{LiPF_6}=1.6$, our simulation reproduces the accelerated degradation rate of LiPF$_6$ very well while preserving the intrinsic SOC dependence derived from the electron diffusion mechanism.

We subsequently examine the influence of the anode material, while keeping the electrolyte salt LiFSI and the temperature $T_\mathrm{ref}=\SI{40}{\degreeCelsius}$ of the standard case. Figure~\ref{fig:capacity-material-choices}(b) depicts the capacity fade for the cells with artificial graphite AGD for cycling in various SOC ranges. Again, cycling in high-SOC regions leads to more SEI growth as compared to cycling in low-SOC regions, in agreement with the trends from the reference case using AGC in \cref{fig:capacity-LiFSI-40}.
For comparison with the reference, we have included the simulation and data for the reference case using AGC for cycling between 0\% and 100\% in grey.
The comparison reveals slightly less capacity loss for AGD in the experiment, while predicting slightly more in our simulation. However, the deviation between simulation and experiment is small due to the overall minor influence of the artificial graphite on the capacity fade at \SI{40}{\degreeCelsius}.

To understand the impact of the artificial graphite and to reason the minor influence on the capacity fade, we have to identify two competing physical mechanisms. First, variations in the anode surface area directly modify the rate of SEI formation in \cref{eq:SEI-model-used}, effectively scaling the prefactor $\xi_\mathrm{AG}$. Second, a larger anode surface area reduces the overall current density, which leads to smaller absolute values of the anode overpotential in \cref{eq:eta} and thus less driving force for SEI growth in \cref{eq:SEI-model-used}. Consequently, these two effects act in opposition regarding the resulting capacity loss. At \SI{40}{\degreeCelsius}, these mechanisms appear to balance out, resulting in very similar degradation profiles for AGC and AGD. Nevertheless, our model captures the influence of the graphite used by scaling both the prefactor and the exchange current density via $\xi_\mathrm{AGD}=2.0$. It is important to note that while the capacity fade profiles are similar at \SI{40}{\degreeCelsius}, we demonstrate in Sec.~\ref{sec:temperatures} that significant differences arise when varying the temperature.

We confirm the consistency of our parameterization by simulations when changing both the electrolyte salt and the graphite at the same time.
The simulation and experimental results for LiPF$_6$ and AGD are shown in the Supporting Information in Fig.~S3(d) and Fig.~S4(d), revealing quite precise agreement. Thus, our model maintains its accuracy when changing the salt and graphite simultaneously, solely using the adjustments for $\xi_\mathrm{electrolyte}$ and $\xi_\mathrm{AG}$ as stated above. 
This consistency demonstrates that our physically derived framework can accurately describe capacity fade across a spectrum of cell material specifications without requiring recalibration for each material combination. After varying material parameters, we will investigate the impact of the operating temperature in the following.

% plot AGC, 40°C, LiPF$_6$ (2x)

\subsection{Influence of operating temperature}
\label{sec:temperatures}

Similarly to the active surface area of graphite, the operating temperature influences SEI growth through two competing mechanisms. First, elevated temperatures accelerate the diffusion of electrons through the SEI layer, leading to faster SEI growth. Second, higher temperatures increase the intercalation kinetics into graphite, which reduces the magnitude of the anode overpotential. Since the driving force for SEI growth is linked to this overpotential, a smaller absolute value results in reduced degradation. 
Hence, the net temperature dependence of SEI growth is non-trivial, as these competing mechanisms interact differently depending on the graphite’s electrochemically active surface area.
Consequently, the effective temperature dependence of SEI growth is intrinsically linked to the properties of the artificial graphite used.

\begin{figure*}[htbp]
    % Figure #
    % width = 2 columns
    \centering
    \includegraphics[height=0.35\textwidth]{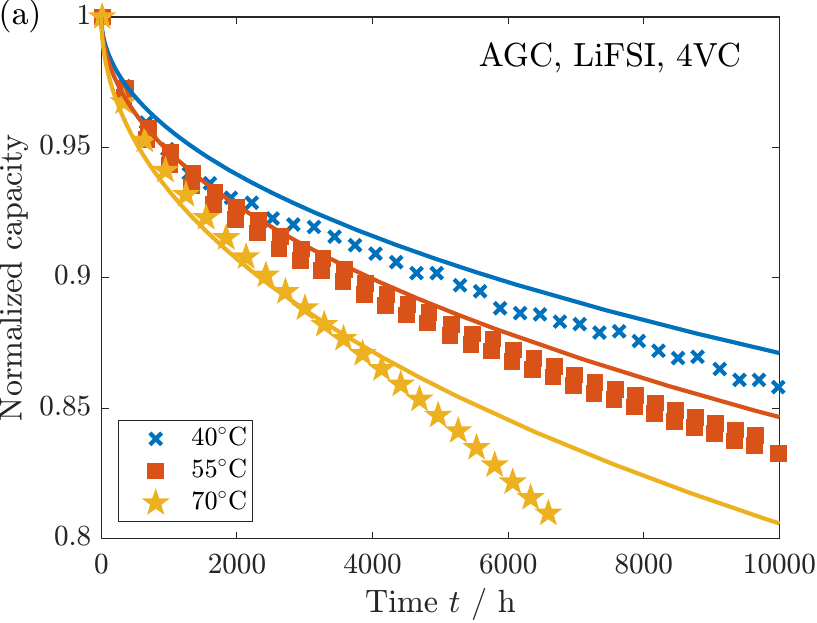}
    \hspace{0.5cm}
    \includegraphics[height=0.35\textwidth]{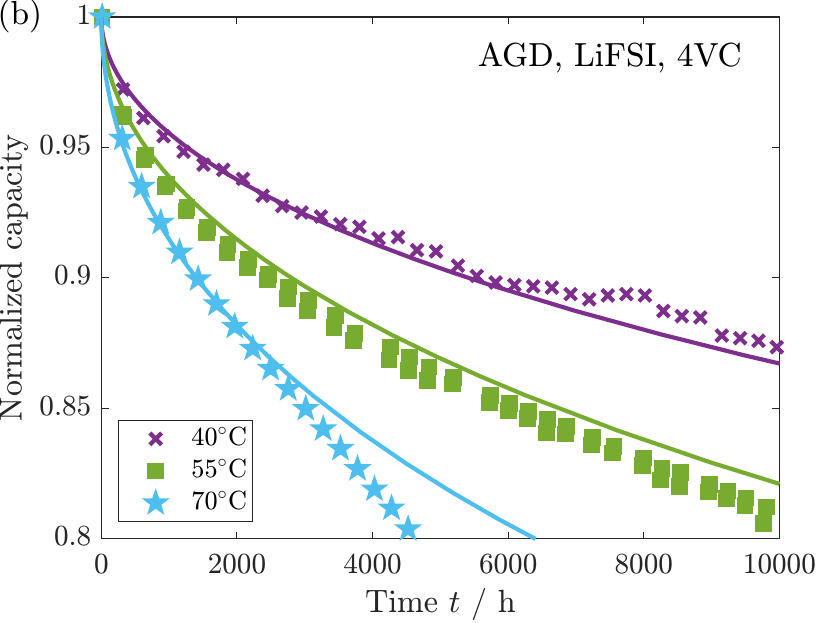}
    \caption{\textbf{Temperature dependence.}
    Evolution of normalized capacity due to SEI growth at \SI{40}{\degreeCelsius}, \SI{55}{\degreeCelsius}, and \SI{70}{\degreeCelsius} for LiFSI and (a) AGC and (b) AGD. Solid lines denote simulation results, while markers indicate experimental measurements.
    Cell IDs: (a) 487692, 487697, 487703. (b) 487672, 487677, 487683.}
    \label{fig:capacity-4VC-Tvariation}
\end{figure*}

Figure~\ref{fig:capacity-4VC-Tvariation} illustrates the capacity evolution of cells cycling between 0\% and 100\% SOC at different temperatures utilizing LiFSI electrolyte with 4\% vinylene carbonate (VC) additive. This additive improves stability and helps maintain square-root time dependence even at elevated temperatures \cite{Black2025}. Figures \ref{fig:capacity-4VC-Tvariation}(a) and (b) show the simulated (lines) and experimental (markers) degradation profiles for AGC and AGD, respectively. While higher temperatures accelerate degradation for both materials, the effect is significantly more pronounced for AGD. Notably, at \SI{70}{\degreeCelsius}, the degradation profile for both graphites deviates from the square-root behavior, indicating the onset of non-linear aging mechanisms, such as those caused by iron dissolution \cite{Black2025}. Similar trends are observed for cells with LiPF$_6$, as detailed in the Supporting Information in Fig.~S5. Cells with 2\% VC additives also show a similar temperature dependence, however, showing non-linear effects at elevated temperatures much earlier as depicted in the Supporting Information in Fig.~S6. Note that AGC might face slight lithium plating at \SI{40}{\degreeCelsius}, increasing the observed capacity fade. However, we assume this effect to be rather negligible compared to the total amount of degradation.

In our model, these temperature dependencies are captured by the activation energy of the electron diffusion, $E_\mathrm{A}$, and the activation energy of the anode overpotential, $E_\mathrm{\eta}$. We determined the activation energy of the electron diffusion, $E_\mathrm{A}$, by fitting the degradation in \cref{fig:capacity-4VC-Tvariation} and the activation energy of the anode overpotential, $E_\mathrm{\eta}$, by fitting the overpotential data specifically for AGD as shown in the Supporting Information in Fig.~S2(a). The difference in the temperature dependence of the two graphites arises because the parameter $\xi_\mathrm{AG}$ changes the impact of the temperature dependence of the electron diffusion step and the temperature dependence of the anode overpotential on the effective SEI growth rate. Here, AGD has a larger active surface area than AGC, leading to a smaller anode overpotential and, consequently, a minor impact of the temperature dependence of the anode overpotential. Thus, the temperature dependence of the electron diffusion step, which increases the degradation with temperature, dominates the temperature dependence of the SEI growth for AGD more strongly than for AGC. This interplay allows our model to accurately reproduce the distinct temperature sensitivities of both graphites, while assuming the same activation energies $E_\mathrm{A}$ and $E_\eta$.

The consistency of our model parametrization is further validated by simulations of experimental cells subjected to varying SOC regions for the different salts and graphites at an elevated temperature of \SI{55}{\degreeCelsius} in the Supporting Information in Fig.~S3(e)-(h) and Fig.~S4(e)-(h). With the complete set of parameters established for all investigated materials and operation temperatures summarized in \cref{tab:fit-parameters}, we have now fully parametrized the model. In the following section, we apply these parameters to our model without further adjustment to investigate the influence of varying cycling rates.

% plot 0-100% cycling at 40, 55, and 70°C

\subsection{Influence of cycling rate}
\label{sec:rates}

Next, we investigate the impact of the cycling rate on the capacity degradation. According to the Butler-Volmer formulation in \cref{eq:eta}, higher applied C-rates directly increase the magnitude of the anode overpotential $\eta$. Since the electron diffusion driving force in \cref{eq:SEI-model-used} scales with $\exp(-F(U_0-\eta)/(RT))$, larger absolute values of $\eta$ at the anode elevate the equilibrium electron concentration at the SEI/electrolyte interface, thereby accelerating SEI formation. Consequently, we anticipate a systematic increase in capacity fade with increasing C-rate, assuming all other conditions remain constant.

\begin{figure*}[htbp]
    % Figure #
    % width = 2 columns
    \centering
    \includegraphics[height=0.35\textwidth]{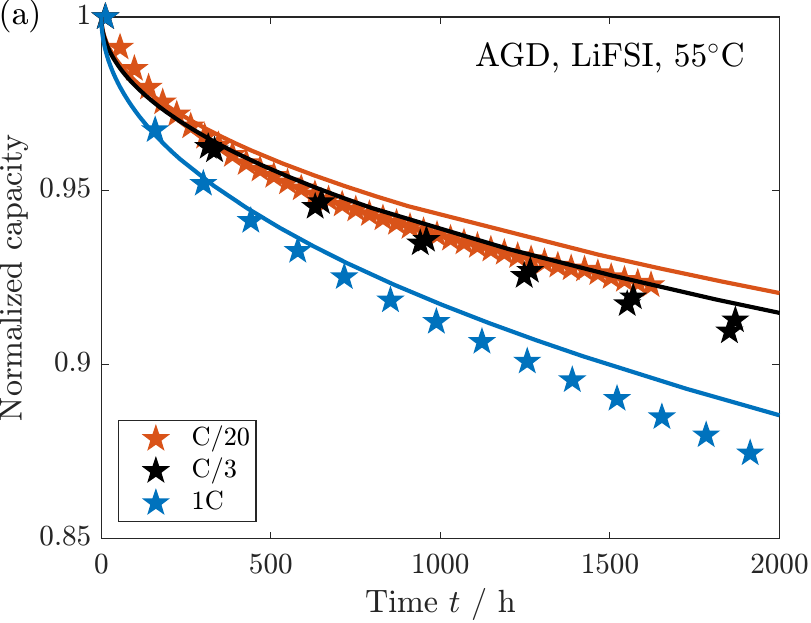}
    \hspace{0.5cm}
    \includegraphics[height=0.35\textwidth]{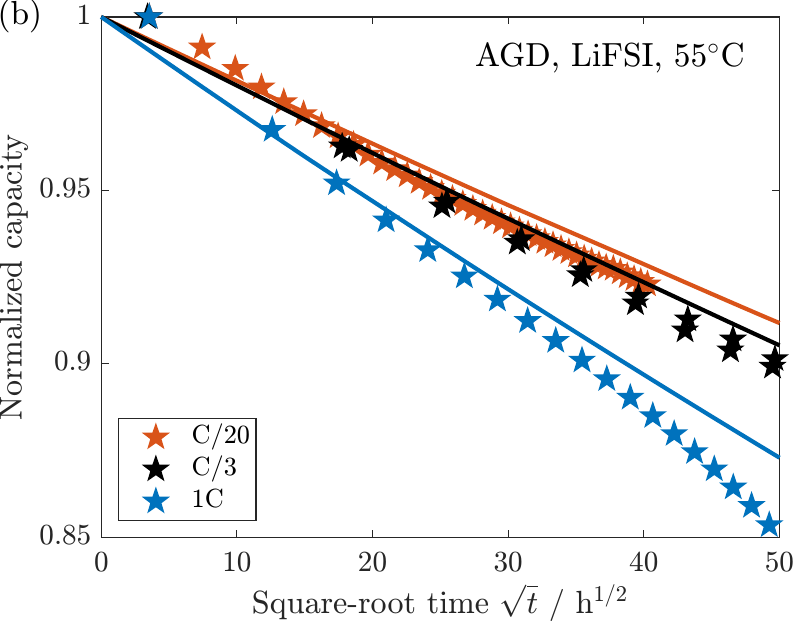}
    \caption{\textbf{Rate dependence.}
    Evolution of normalized capacity due to SEI growth for AGD and LiFSI at \SI{55}{\degreeCelsius} for different C rates. (a)  Capacity evolution as a function of time. (b) Capacity evolution plotted versus the square root of time. 
    Solid lines denote simulation results, while markers indicate experimental measurements.
    Cell IDs: 1076296, 1075229, 487677.}
    \label{fig:capacity-LiFSI-55-AGD-rates}
\end{figure*}

This expectation is experimentally validated by the degradation profiles of AGD cells with LiFSI electrolyte with 4\% VC additive cycled at \SI{55}{\degreeCelsius} with 1C, C/3, and C/20 rates, as shown in \cref{fig:capacity-LiFSI-55-AGD-rates}(a). The graphite AGD and the elevated temperature were selected for this test to minimize the risk of lithium plating. As anticipated, higher C-rates induce a larger capacity loss in the experimental data (markers). Our simulations (lines), which incorporate the C-rate-dependent overpotential from \cref{eq:eta}, reproduce this trend with high fidelity, confirming that the observed acceleration in degradation arises directly from the higher overpotential at higher rates. When plotted against the square root of time in \cref{fig:capacity-LiFSI-55-AGD-rates}(b), the capacity evolutions for the different cycling rates exhibit a linear trend, demonstrating the square-root-in-time degradation profile typical of diffusion-limited SEI growth. Minor deviations from linearity appear only in the 1C dataset, marking the onset of additional degradation mechanisms such as iron dissolution under these accelerated conditions \cite{Black2025}.

The influence of the C-rate is further explored at elevated temperature in the Supporting Information, where Fig.~S7 presents the 1C, C/3, and C/20 profiles at \SI{70}{\degreeCelsius}. The fundamental trend that higher currents yield faster degradation remains consistent, and the simulation overall captures this behavior. However, consistent with the \SI{55}{\degreeCelsius} data, the 1C curve at \SI{70}{\degreeCelsius} exhibits even more pronounced non-linear degradation, which aligns with expectations for an increased impact of iron dissolution at \SI{70}{\degreeCelsius} \cite{Black2025}.

These results demonstrate that our SEI growth model based on the electron diffusion mechanism successfully captures the C-rate dependence of capacity fade during cycling, solely through the physical coupling of current density, anode overpotential, and the SEI growth rate. Next, we turn to protocols that intermix cycling with open-circuit storage periods to evaluate the model's ability to cope with mixed aging conditions.

\subsection{Influence of mixed cycling and storage protocols}
\label{sec:storage-periods}

Beyond continuous cycling, we now investigate the model's performance under protocols that interleave cycling with open-circuit storage periods. During storage, the anode overpotential $\eta$ vanishes, and the degradation rate is governed strictly by the open-circuit voltage, $U_0(\text{SOC})$, which determines the driving force for the electron diffusion mechanism.

\begin{figure*}[htbp]
    % Figure #
    % width = 2 columns
    \centering
    \includegraphics[height=0.35\textwidth]{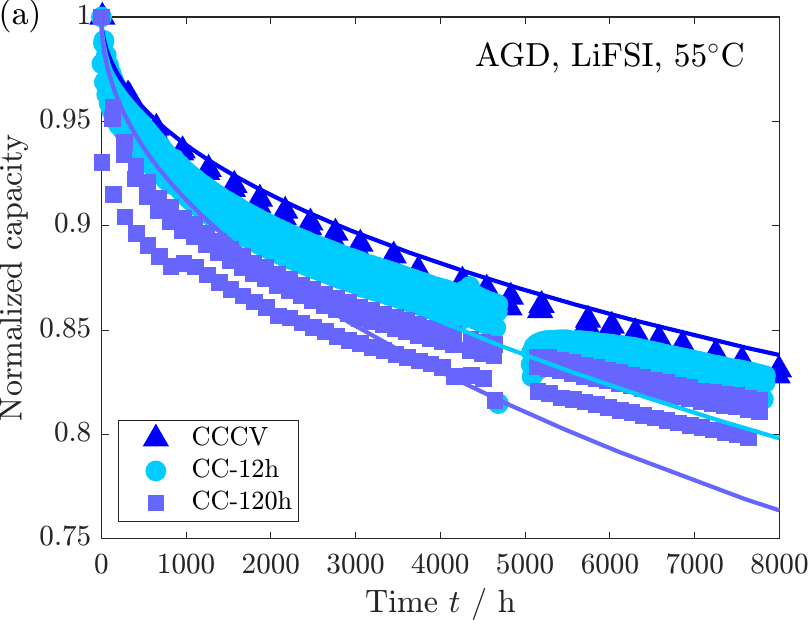}
    \hspace{0.5cm}
    \includegraphics[height=0.35\textwidth]{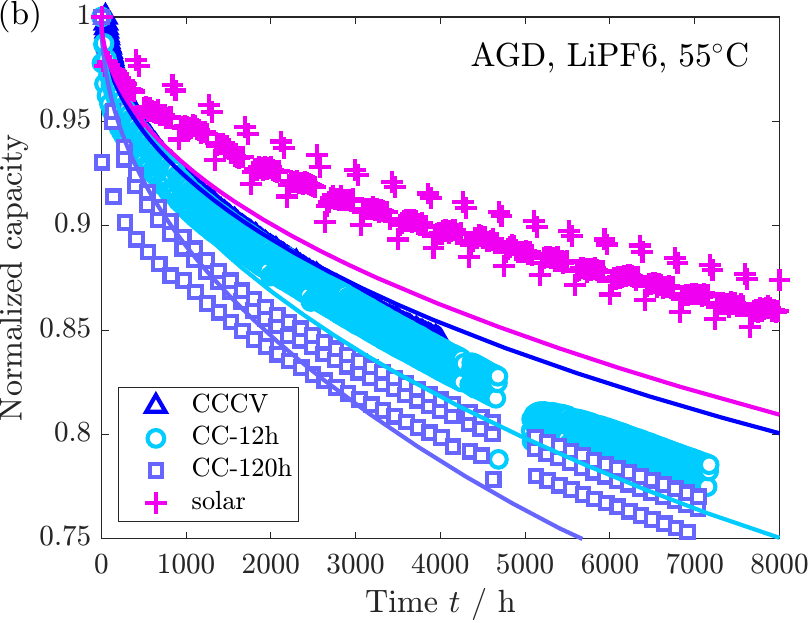}
    \caption{\textbf{Protocol dependence.}
    Evolution of normalized capacity due to SEI growth for AGD at \SI{55}{\degreeCelsius} for different protocols for (a) LiFSI and (b) LiPF$_6$ electrolyte salt. 
    Solid lines denote simulation results, while markers indicate experimental measurements. 
    Cell IDs: (a) 487677, 723110, 723119. (b) 967710, 723107, 723117, 721536.}
    \label{fig:capacity-protocols}
\end{figure*}

Figure \ref{fig:capacity-protocols} depicts the capacity fade evolution at \SI{55}{\degreeCelsius} for AGD cells using LiFSI electrolyte in \cref{fig:capacity-protocols}(a) and LiPF$_6$ electrolyte in \cref{fig:capacity-protocols}(b), each with 4\% VC additive. The experiments compare pure 0\%--100\% SOC cycling against protocols that include storage periods. Specifically, we examine the addition of \SI{12}{h} and \SI{120}{h} storage periods at 100\% SOC, as well as a solar duty cycle protocol that introduces slower charging and storage periods at both 0\% and 100\% SOC as detailed in Sec. \ref{sec:experiments}.

The experimental data, shown as markers in \cref{fig:capacity-protocols}, reveal that inserting storage periods at 100\% SOC generally accelerates the initial capacity loss compared to pure cycling, while the degradation rate appears to slow down at later stages. The deviation from the typical square-root-of-time dependence is also evident when plotting versus square-root time in the Supporting Information in Fig.~S8. The reason for this deviation is not clear but suggests a passivation effect, wherein the SEI layer becomes more protective over time during extended rest. This exact mechanism would need further investigation and is beyond the scope of our study. Moreover, as the cycle protocols with storage at 100\% do not have regular checkups, the degradation is calculated with the C/3 capacity instead of the C/20 capacity, which is unfortunately expected to influence the profile.
In contrast to the protocols with extended storage periods at 100\% SOC, the solar duty cycle protocol, which particularly includes storage periods at 0\% SOC, results in significantly reduced capacity loss compared to the other protocols. This confirms the strong SOC sensitivity of SEI growth, as the low-potential state at 0\% SOC during storage minimizes the driving force for SEI growth.

The simulation results, shown as solid lines in \cref{fig:capacity-protocols}, qualitatively reproduce these trends. The model correctly predicts that 100\% SOC storage accelerates the degradation and that the solar duty cycle results in the least severe degradation. However, the model quantitatively overestimates the capacity loss for all protocols involving storage at 100\% SOC, with the solar duty cycle showing the largest discrepancy. We believe this overestimation is primarily driven by an excessive prediction of degradation during the 100\% SOC rest periods.

The reason for the overestimation of capacity fade during high-SOC storage is not clear. Literature studies on high-voltage cathodes like nickel-manganese-cobalt-oxide have reported reduced capacity loss during storage at 100\% SOC compared to slightly lower states (e.g., 80–90\% SOC), potentially due to pronounced electrolyte oxidation at the cathode \cite{Keil2017, Zuelke2021, Wildfeuer2023}. However, this phenomenon is not expected for low-voltage LFP cathodes, although some studies also find slightly reduced capacity fade at 100\% SOC as well \cite{Keil2016, Naumann2018}. 
In addition, some uncertainty remains in the precise estimation of the anode voltage in full cells.
Importantly, the excellent agreement during slow cycling with C/20 in \cref{fig:capacity-LiFSI-55-AGD-rates} suggests that the model’s fundamental degradation kinetics are sound, and the deviation is likely specific to the unique electrochemical environment of high-SOC storage rather than a general difference in storage and cycling degradation mechanisms.

The mixed cycling and storage protocols show that our simulations capture the general trends of the degradation for intermixed cycling and storage periods, although these trends are not quantitatively captured, possibly due to the limitations discussed above. Finally, we evaluate the overall quality of our simulations by comparing predicted versus observed degradation for all presented data in the following section.

% plots AGD, LiPF$_6$ and LiFSI at 55°C

\subsection{Global model validation}
\label{sec:consistency}

To demonstrate the consistency of our physical SEI growth model across the entire experimental dataset, we perform a quantitative comparison between the simulated and observed amounts of capacity loss. This assessment encompasses all cycling and storage protocols, electrolyte salts, artificial graphites, and operating temperatures investigated in both the main manuscript and the Supporting Information. We calculate the Root Mean Square Deviation (RMSD) to provide a statistical measure of the agreement across this broad range of conditions.

\begin{figure*}[htbp]
    % Figure #
    % width = 2 columns
    \centering
    \includegraphics[height=0.35\textwidth]{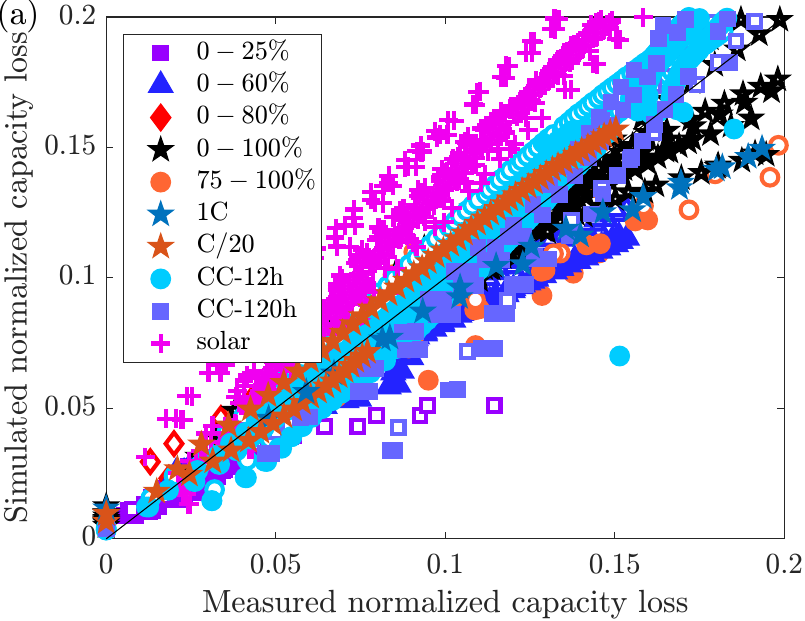}
    \hspace{0.5cm}
    \includegraphics[height=0.35\textwidth]{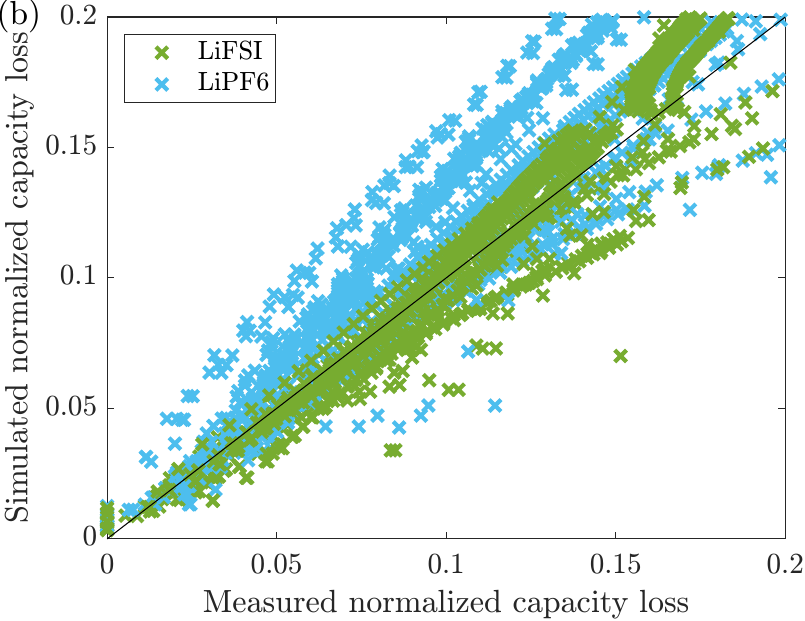}
    
    \includegraphics[height=0.35\textwidth]{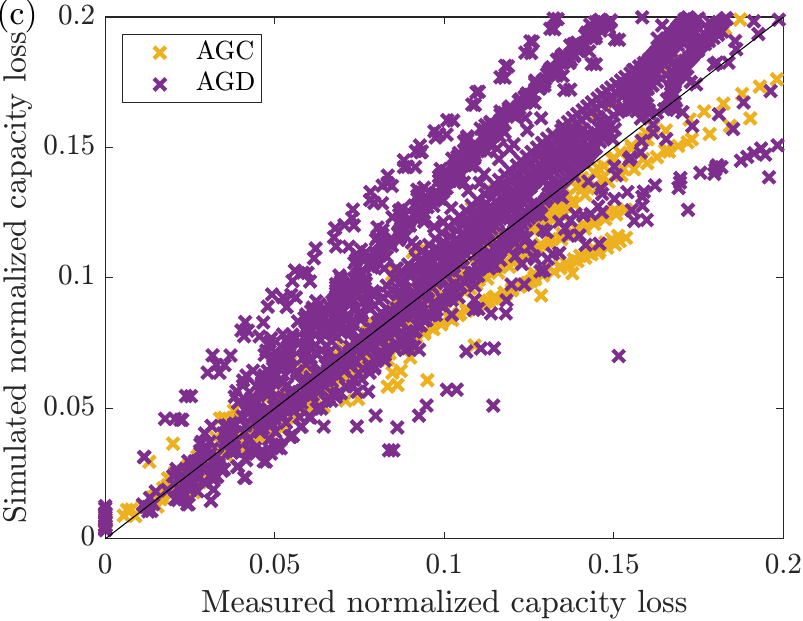}
    \hspace{0.5cm}
    \includegraphics[height=0.35\textwidth]{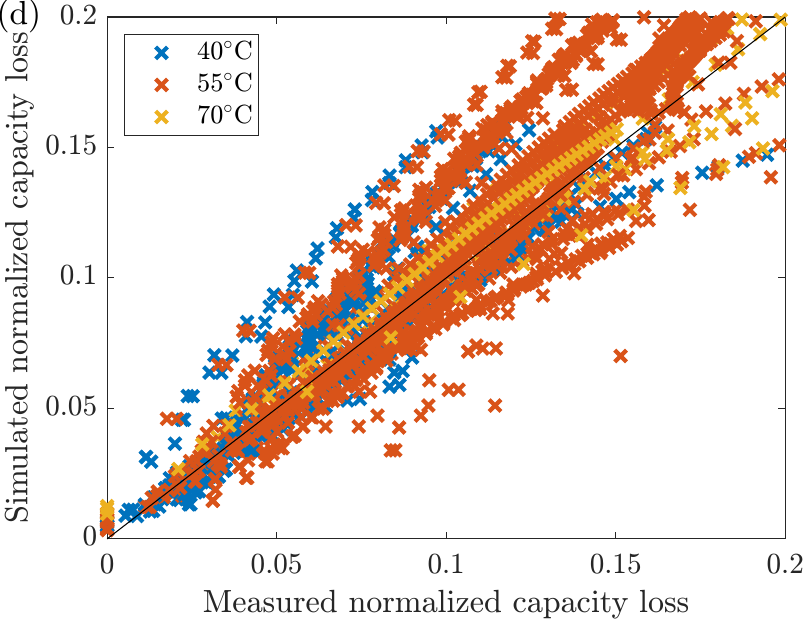}
    \caption{\textbf{Model validation.}
    Simulated vs observed normalized capacity loss due to SEI growth for all cells investigated in the manuscript and the Supporting Information. All subplots show the same data points, but color code the different cycling conditions or cell specifications: (a) Cycling protocol (SOC window, C-rate, and mixed storage and cycling protocol), (b) electrolyte salt, (c) artificial graphite, and (d) operating temperature.}
    \label{fig:capacity-overview}
\end{figure*}

We plot the simulated capacity loss against the observed capacity loss in \cref{fig:capacity-overview}. The data points follow a linear trend with high correlation. We calculate the RMSD $\Delta$ of the normalized simulated capacity loss $\tilde{Q}_\mathrm{SEI}^\mathrm{sim}$ and the normalized experimental capacity loss $\tilde{Q}_\mathrm{SEI}^\mathrm{exp}$ according to
\begin{equation}
    \Delta = \sqrt{\frac{1}{N} \sum \left(\tilde{Q}_\mathrm{SEI}^\mathrm{sim} - \tilde{Q}_\mathrm{SEI}^\mathrm{exp}\right)^2}
\end{equation}
with the number of data points $N$. The comparison of simulation and experiment yields a global RMSD of $\Delta=2.7\%$. To provide a detailed inspection of the model's performance across different parameters, the figure is divided into subplots that re-plot the same data using distinct color coding to highlight specific subsets. 
Figure~\ref{fig:capacity-overview}(a) focuses on the cycling protocols. As discussed in Sec.~\ref{sec:storage-periods}, the strongest deviations are observed for the solar duty cycle and protocols incorporating storage periods. These discrepancies are primarily attributed to the model's overestimation of degradation during 100\% SOC storage. Excluding these cases, the deviations for the different cycling protocols predominantly appear only at later stages of aging, corresponding to a better RMSD of $\Delta=1.8\%$. In addition to the protocols,
\cref{fig:capacity-overview}(b) highlights the variation in electrolyte salt, \cref{fig:capacity-overview}(c) focuses on the artificial graphite, and \cref{fig:capacity-overview}(d) emphasizes the temperature effects. With the exception of the outliers associated with the solar duty cycle, the plots demonstrate a good agreement between simulation and experiment across all salts, graphites, and temperatures.

In summary, these results demonstrate the quality of our physically derived SEI growth model based on the electron diffusion mechanisms in reproducing a wide spectrum of experimental aging data.  The agreement is achieved using a single set of parameters that accounts for the influence of salt, graphite, cycling protocol, and temperature as summarized in \cref{tab:fit-parameters} without the need for recalibration. Most importantly, the SOC-dependence of the degradation is captured intrinsically by the foundations of the electron diffusion mechanism. While limitations remain for protocols containing storage periods, the model successfully describes the fundamental degradation mechanism under diverse conditions, validating its predictive capability across a broad range of cell specifications and operating conditions.

% overview plot to show consistency

\section{Conclusion}
\label{sec:conclusion}

In this work, we have presented and validated a physically grounded SEI growth model based on the electron diffusion mechanism, adjusted to reproduce the degradation observed across different cell specifications. By systematically comparing simulation results against a comprehensive set of experimental aging data, we demonstrated that this framework provides a consistent description of the capacity degradation across a broad spectrum of cell specifications and operating conditions.

A key finding of this study is that the model intrinsically captures the experimentally observed SOC dependence of SEI growth during cycling. For cycling in confined SOC ranges, the electron diffusion model reproduces the capacity fade without the need for empirical tuning, confirming that the anode potential predominantly governs the degradation rate. Furthermore, we established the high transferability of the model: a consistent set of parameters successfully predicts aging behavior across various electrolyte salts (LiFSI, LiPF$_6$), types of artificial graphite (AGC, AGD), operating temperatures (\SI{40}{\degreeCelsius}, \SI{55}{\degreeCelsius}, \SI{70}{\degreeCelsius}), and C-rates (C/3, 1C, C/20). This consistency underscores the model's potential as a predictive tool for diverse battery specifications and cycling conditions.

While the model performs very well under various cycling conditions, challenges remain when addressing elevated temperatures as well as mixed cycling and storage protocols with storage periods at high SOC. 
At elevated temperatures, non-linear degradation effects beyond the scope of our model become more prominent, particularly the impact of iron dissolution \cite{Black2025}.
For protocols involving storage periods at high SOC, we observed discrepancies between simulations and experiments. The reason for these discrepancies is not clear. However, some uncertainty remains in the precise estimation of the anode voltage in full cells. Consequently, future improvements in modeling accuracy will likely depend on more precise experimental determination of the exact anode open-circuit voltage and anode overpotential present in full cells.

Despite these limitations, this work contributes significantly to the fundamental understanding of LFP/graphite lithium-ion battery degradation. The validated model demonstrates that operating and storing cells at lower SOC regions is an effective strategy to mitigate degradation and extend cell lifetime. Looking forward, to achieve even longer cell lifespans, research efforts should focus on reducing the SEI growth prefactor by developing strategies that more effectively impede electron diffusion through the SEI layer. Our physically derived framework provides a solid basis for evaluating such material advancements in future battery designs.

\section{Acknowledgments}

We thank Kübra Senel for performing the half-cell experiments.
Lukas Köbbing gratefully acknowledges funding and support from the European Union’s Horizon Europe within the research initiative Battery 2030+ under the grant agreement No. 101104032 (OPINCHARGE) and No. 101103997 (DigiBatt). This work contributes to the research performed at CELEST (Center for Electrochemical Energy Storage Ulm-Karlsruhe).
The Dalhousie authors acknowledge funding from NSERC and Tesla Canada under the auspices of the Alliance Grants program.  Claire Floras and William Black acknowledge NSERC for scholarship support.  Claire Floras thanks the Tyler Lewis Clean Energy Research Foundation and the Mavroyannis Family Grant for support.

\section{Declaration of Conflict of Interest}
The authors declare no conflict of interest.

\section{CRediT Authorship Contribution Statement}

\textbf{Lukas Köbbing}:
Conceptualization,
Methodology,
Software,
Validation,
Formal analysis,
Investigation,
Data Curation,
Writing - Original Draft,
Visualization.
\textbf{William Black}:
Conceptualization,
Validation,
Investigation,
Data Curation,
Writing - Review \& Editing.
\textbf{Claire Floras}:
Conceptualization,
Validation,
Investigation,
Data Curation,
Writing - Review \& Editing.
\textbf{Saad Azam}:
Conceptualization,
Validation,
Investigation,
Data Curation,
Writing - Review \& Editing.
\textbf{Eniko Zsoldos}:
Conceptualization,
Validation,
Investigation,
Data Curation,
Writing - Review \& Editing.
\textbf{Ben Tang}:
Conceptualization,
Validation,
Investigation,
Data Curation,
Writing - Review \& Editing.
\textbf{J.R. Dahn}:
Conceptualization,
Methodology,
Resources,
Writing - Review \& Editing,
Supervision,
Project administration,
Funding acquisition.
\textbf{Birger Horstmann}:
Conceptualization,
Methodology,
Resources,
Writing - Review \& Editing,
Supervision,
Project administration,
Funding acquisition.

\bibliography{refs}% Produces the bibliography via BibTeX.

\end{document}